\documentclass[sn-mathphys-num]{sn-jnl}

\usepackage{setspace}
\usepackage{graphicx}%
\usepackage{multirow}%
\usepackage{amsmath,amssymb,amsfonts}%
\usepackage{amsthm}%
\usepackage{mathrsfs}%
\usepackage[title]{appendix}%
\usepackage{xcolor}%
\usepackage{textcomp}%
\usepackage{manyfoot}%
\usepackage{booktabs}%
\usepackage{algorithm}%
\usepackage{algorithmicx}%
\usepackage{algpseudocode}%
\usepackage{listings}%

\theoremstyle{thmstyleone}%
\theoremstyle{thmstyletwo}%

\theoremstyle{thmstylethree}%

\begin{document}

\title[Deriving Ontological Statements from the Unnatural Higgs Mass]{Deriving Ontological Statements from the Unnatural Higgs Mass}

\author{\fnm{Johannes} \sur{Branahl (j\_bran33@uni-muenster.de)}}

\affil{\orgdiv{Philosophisches Seminar}, \orgname{Universit\"at M\"unster}, \orgaddress{\street{Domplatz 23}, \city{48143 Münster},  \country{Germany}}}

\abstract{We provide novel, metatheoretical arguments strengthening the position that the naturalness problem of the light Higgs mass is a pseudo-problem: Under one assumption, no physics beyond the standard model of particle physics is needed to explain the small value of the Higgs boson. By evaluating previous successes of the guiding principle of technical naturalness, we restrict its applicability to non-fundamental phenomena in the realm of provisional theories within limited energy scales.  In view of further breaches of autonomy of scales in apparently fundamental phenomena outside particle physics, the hierarchy problem of the Higgs mass is instead reinterpreted as an indication of the ontologically fundamental status of the Higgs boson. Applying the concept of robustness of theoretical elements under theory changes by Worrall and Williams justifies this seemingly contradictory attribution within the effective theories of the standard model of particle physics.  Moreover, we argue that the ongoing naturalness debate about the Higgs mass is partly based on the adherence to the methodology of effective theories (often claimed to be universally applicable), for which there is no justification when dealing with presumably fundamental phenomena such as the Higgs mechanism, even if it is embedded into an effective theory.}
\maketitle

\subsection*{Conflict of interests/funding}
The author states that there is no conflict of interest. He did not receive support from any organization for the submitted work.
 \section{Introduction}

This article deals with the naturalness problem of the Higgs mass, also known as the hierarchy problem. Despite numerous attempts to characterize the problem, to question its status as a problem, or to address it with new physics: One is far from a consensus, especially since, in some interpretations, it questions the theoretical framework of particle physics and has profound philosophical implications. In particular, the past decade has seen a fruitful exchange between physicists and philosophers of science, which this work seeks to join.

The naturalness principle, independent of its precise definition, regards the physicist's expectation of numerical values for parameters. By a majority that still shapes the research program, theories in particle physics are still considered unnatural and in need of explanations, if such numerical values turn out to be particularly small or large in experiments. Naturalness is percepted as the central concept "if one has to summarize in one word what drove the efforts in physics beyond the standard model of the last several decades" (Giudice 2017, p.\,3). Naturalness has become a core criterion in theory assessment in the standard model of particle physics as well as physics beyond the standard model (BSM physics for short), and its violation has been and is still considered as something in need for an explanation.\;It may seem surprising that the efforts of BSM physics have been significantly driven for half a century by such an inconspicuous criterion as the order of magnitude of certain constants of nature. Hence, given that significant progress in the history of physics has been made due to unexplainable empirical observations or theoretical inconsistencies (which both do not occur in case of unnatural parameters), naturalness requires a critical revision of its validity and relevance as a criterion for theory evaluation. 

An upfront concession: In the virtual absence of unexplainable empirical results in the BSM physics research program (except of neutrino oscillations and possibly dark matter) with, simultaneously, some plausibility for BSM physics at the yet unknown energy scales up to the Planck energy, resorting to guiding principles like naturalness seems initially legitimate. In a highly developed research program where fundamental limits of the technically possible are imposed in the search for empirical evidence, astonishment about certain numerical values should also be paid attention to. This is especially the case when unnatural values impact cornerstones of theory-building in this research program: symmetries and effective field theory, namely in the definitions of 't Hooft naturalness (1979) and technical naturalness (Susskind 1979). These concepts attribute unnaturalness in two important cases - to the $\bar \theta$ parameter of QCD (strong CP problem) and the Higgs mass in the standard model. Both might go beyond mere aesthetic discomfort. However, numerous attempts to solve both problems with BSM physics never led to empirical progress. Additionally, the debate on technical naturalness becomes even more significant with its interdisciplinary emergence, as it also arises in the context of the cosmological constant\footnote{See (Craig 2022) for an overview of the strong CP problem, the hierarchy problem and the cosmological constant.}. 

This article provides novel (metatheoretical) arguments that the restoration of technical naturalness of the Higgs mass with new physics represents an engagement with a pseudo-problem, thus hindering the progress of the BSM physics program. We will discuss reasons that there is no need for a physical explanation of a violation of technical naturalness in this particular case. This critique of the presumably most crucial non-empirical guiding principle of BSM physics, which significantly contributed to the multibillion-dollar construction of the LHC in Geneva\footnote{Of course, confirming the hypothetical Higgs boson itself was perhaps an even stronger reason to build the LHC. However, this discovery was not its only purpose.}, is not isolated. In recent times, several calls for a re-evaluation of naturalness have been made. Prominent critics include the CERN theorist Giudice (2017) and the author, physicist, and philosopher Hossenfelder (among others: 2018, 2021). While Giudice, formerly a staunch defender of naturalness, offers mild criticism and proclaims a post-naturalness era but wishes to retain naturalness as a still significant criterion, Hossenfelder is uncompromising. She calls for a complete departure from the principle of naturalness, which she believes has misled particle physics for too long\footnote{Other contributions to the naturalness debate include (Giudice 2008, Wells 2013 Dine 2015, Hossenfelder 2017, Williams 2015, 2019, Borrelli and Castellani 2019, Wallace 2019, Bain 2019, Harlander and Rosaler 2019, Franklin 2020, Koren 2020, Rosaler 2021, Athanasiou 2022, Fischer 2023, 2024). Our apologies go to all authors whose contributions were not mentioned.}.

In this article, at least some significance - in contrast to Hossenfelder - should be attributed to the small value of the Higgs mass, not creating a need for explanation, but indicating that the Higgs boson is ontologically fundamental and placing it outside the provisional, effective concepts of the standard model physics (effective field theories, EFT for short). The unnatural Higgs mass does not pose a problem for high-energy physics, but instead allows for ontological commitments about high-energy physics. The breakdown of autonomy of scales marks the end of a pragmatic approach in particle physics that cannot meet the claim of fundamental physics to make definitive statements about the world - ultimately, the desire to relieve philosophy of the task of ontology. Instead of being horrified by the failure of effective theories, a sensitivity to large energy scales should provide ample reason for optimism that the overarching goal of fundamental physics will succeed. 

 In summary, two consecutive hypotheses will be developed. It is a key element of this article to clarify the distinction between technical naturalness as a principle that applies to EFTs and the idea that some entities within EFTs might be ontologically fundamental despite the provisional nature of the overall theory. The vocabulary of ontology instead of EFTs helps distinguish between phenomena that are genuinely provisional and those that might survive in any future theory. This new wording resharpens the common expression for the limited applicability of technical naturalness: \\ \\
$H_1$: \textit{The metatheoretical guiding principle of technical applies strictly to ontologically non-fundamental phenomena modelled in the provisional framework of EFTs.} \\ \\ 
 This revision sheds new light on the potentially misleading deduction: "The standard model is an EFT, and the Higgs boson is embedded into the standard model, so the principle of technical naturalness needs to apply - if not, a hierarchy problem arises". Instead, it turns around the entire argumentation of the hierarchy problem, because the Higgs boson operates beyond the confines of provisional EFTs: \\ \\ 
$H_2$: \textit{The light Higgs mass is rather an indication of its fundamental ontological status and its special role outside of effective theories.} \\ \\ 
As a result, the so-called naturalness problem associated with the Higgs mass is better understood as a pseudo-problem, arising from the misapplication of naturalness principles that are not relevant to ontologically fundamental phenomena. The argumentation for these hypotheses will proceed as follows: In Chapter 2, we first recapitulate some important definitions of naturalness and particularly describe the lack of technical naturalness of the renormalized Higgs mass. Confronted with this problem, there are three options: firstly, attempting to address it with new physics but repeatedly facing disappointment due to the experimental failures of all previous theories. Secondly, attempting to completely deny the problem: the Higgs mass is a constant of nature, and discussing its value becomes unnecessary. We opt for the third option and question the adherence to effective field theories. To do so, we argue in Chapter 3 that quantum corrections to the Higgs mass can indeed be ascribed a physical reality, and the dependence of the Higgs mass on high-energy domains has significance for theory evaluation: the breakdown of autonomy of scales. However, we then narrow down the scope of technical naturalness by explaining several phenomena in which its violation occurs. For each of these phenomena, a fundamental ontological status can be attributed, and the violation of autonomy of scales does not pose a problem ($H_1$). In Chapter 4, we argue against the dogmatic adherence to this principle and thus to effective theories as the only way to do physics, as the philosophical consequences of this adherence are incoherent with the actual modus operandi of physics and the worldview revealed therein. 

Finally, we attempt to reconcile the fundamental nature of the Higgs mass with the non-fundamental character of an EFT (the standard model as a whole) diagnosing a pseudo-problem of the small Higgs mass by means of $H_2$. Hence, we derive a positive reaction to the violated autonomy of scales in case of the Higgs boson - that humankind uncovered a fundamental entity and did not created a provisional construction for preliminary empirical adequacy in limited energy scales and experimental resolutions.  In the conclusion of Chapter 5, the significant role of effective theories for a thriving future particle physics is emphasized. Nonetheless, occasional detachment from their principles is necessary to avoid pseudo-problems like that of the unnatural Higgs mass.

\section{Technical Naturalness and the Hierarchy Problem}
We start with a recapitulation of naturalness in general, technical naturalness in particular, and its connection to the Higgs mass. Apart from definitions of weak and strong autonomy of scales, the reader familiar with the literature on the hierarchy problem may directly jump to Chapter 3.

\subsection{Definitions of Naturalness} 
In physics in general, the concept of naturalness arises in the context of Dirac's Large Number Hypothesis: Given an operator $\xi$ of the form $\mathcal{L} \propto c_0 \xi$, where $\mathcal{L}$ represents the Lagrangian of the physical theory, the dimensionless parameter $c_0$ in natural units is expected to be of the order $\mathcal{O}(1)$ (Dirac 1974). Throughout the last century, Dirac repeatedly appears as an advocate of aesthetic criteria as a decisive basis for evaluation in physics. Similarly, this notion does not have underlying physical principles justifying a dogmatic adherence to absolute naturalness (cf.\;Wells 2013). It appears to be solely based on a desire for numerical aesthetics. Although the principle of absolute naturalness often proves correct, it fails for many parameters of the standard model and must, therefore, be considered too broad and unsuitable for further discussion.

In the context of the standard model, to the best of our knowledge, Weinberg (1972) first mentions naturalness in an attempt to find a "natural explanation of the approximate symmetries in nature". Symmetry, whose profound significance for particle physics is undoubted, can restrict the demand for absolute naturalness. This 't Hooft naturalness (1979) allows $c_0$ to take a much smaller value than $\mathcal{O}(1)$ as long as the limit $c_0 \to 0$ leads to a symmetry. Chiral symmetries or gauge symmetries, for example, protect the fermion masses or gauge boson masses of the standard model from (unnatural) large correction terms in interaction with virtual particles, justifying the constraint imposed by 't Hooft. Mathematically, this cannot happen for the Higgs boson since this particle, being the only one, has no spin - its mass is 't Hooft-unnatural. The definition of naturalness that is more relevant for this chapter is reproduced from the 't Hooftian perspective but further narrows down the set of unnatural phenomena. In this work, this is defined as technical naturalness\footnote{In the literature, this term is sometimes used synonymously with 't Hooft naturalness. Instead, in the present context, the term \textit{AoS} naturalness (autonomy of scales) is used. However, the unfortunate term \textit{technical naturalness}, sounding somewhat contradictory, cannot be avoided even in this convention.}. It describes the prohibition of mutual influence from energy scales separated by many orders of magnitude (autonomy of scales) and first appears in Susskind (1979). We will sharpen that definition after having provided somewhat more background\footnote{Readers interested in further details about the history of the naturalness debate will find a very readable introduction in (Borrelli, Castellani 2019).}.

Symmetries, in the spirit of 't Hooft naturalness, can preserve technical naturalness, as seen previously with the masses of the SM. However, the small $\bar \theta$ parameter of QCD does not violate autonomy of scales but is 't-Hooft-unnatural. For this reason, the violation of technical naturalness must already be considered a clearer indication of a need for explanation, as it implies 't-Hooft unnaturalness (the absence of a protective symmetry against a correspondence between widely distant scales) - at least if naturalness is considered a valid guide in general. Furthermore, the demand for autonomy of scales is closely related to the construction principle of an effective field theory (EFT), which, in the standard model, uses the cutoff parameter in the renormalization of particle masses. In contrast, the desire for approximate symmetries when finding small parameters according to 't Hooft appears more as a straightening of the aesthetics of the theory: If one must accept unsightly parameters, they should at least come with a symmetry. Criticism of the present modus operandi of BSM physics, accessible to a broader readership in content, occasionally suggests that physicists are more guided by aesthetic requirements for parameters than the genuinely physically, substantial definition of technical naturalness reveals. This is not to deny the actual importance of aesthetics in the search for BSM physics but to clarify the theory-internal significance of technical naturalness and separate it from the aesthetics of science. It sometimes conveys a misleading impression in the scientific literature when both the strong CP problem and the hierarchy problem of the Higgs mass are mentioned in one sentence under the umbrella term of naturalness (see, for example, Hossenfelder 2018, Giudice 2017, Craig 2022). These are two very different concepts.

\subsection{Renormalization of the Higgs Mass}
Determining the absolute strength of fundamental interactions in the Universe, one is confronted with a very clear hierarchy (giving the naturalness problem its name, as explained now). The extremely weak effect of gravity compared to all other forces, with a factor of $10^{-41}$ compared to the strength of electromagnetic interaction on the quark length scale, is initially fortunate: Gravity only acts as a dominant interaction in the case of large mass accumulations (such as planets and stars), whereas in the subatomic realm its effect can be neglected and no quantum gravity is needed yet.

Unfortunately, gravity returns through the back door into quantum field theory as we venture closer to the Big Bang, involving higher temperatures of the universe, and thus higher particle energies. On this journey, one traverses "the desert", the unknown physics that can extend over 15 orders of magnitude up to the Planck scale.  Due to the lack of a theory of quantum gravity, precise statements about processes at the maximum of this scale cannot be made. Therefore, in quantum field theory, each integral over particle energies is cut off where new physics is expected, and quantum field theory is superseded - with the cutoff $\Lambda \propto M_{\mathrm{Planck}}$ (or already at $M_{GUT}$). The standard model becomes an EFT.

 The procedure of renormalization removes the cutoff-dependence: In the initial formulation of renormalization during the development of QED, the limit $\Lambda \to \infty$ was considered. This formalistic idea of renormalization, which caused discomfort to physicists of this generation precisely because of its infinities (cf.\,Kragh 1990, p.\,1984 on Dirac), fails due to the fundamental limits of the applicability of QFT (quantum gravity, Landau poles). Wilson's (1975) significant contribution, a representative of the next generation of quantum field theorists, was ultimately based on the reinterpretation of formalistic renormalization as realistic renormalization with a finite cutoff. He conceives QFT as a limited, effective field theory and exemplifies this idea with the requirement $\Lambda \propto M_{\mathrm{Planck}}$. This contemporary understanding of renormalization reveals that QFT cannot be the ultimate answer to questions in fundamental physics. However, the finite cutoff of the renormalizable theory is finally the reason "that we can do sensible macroscopic physics even without having detailed knowledge of the underlying microscopic theories" (Adlam 2019, p.\,10).

With these preparatory remarks, we can turn to the Higgs mass: After renormalization, the parameter of the bare Higgs mass $m_{H,b}$ is replaced by the renormalized Higgs mass $m_{H,r}$, into which the perturbative corrections (self-energy) $\Delta m_{H}$ went. More precisely: $m_{H,r} = m_{H,b} + \Delta m_H$. The calculation of fermionic loops in first order yields the following correction:
\begin{align*}
\Delta m_H^2 = \frac{\Lambda^2}{16 \pi^2 } \biggl (-6y_t^2 + \frac{9}{4}g^2 + \frac{3}{4} g'^2 + 6 \lambda \biggl )
\end{align*}
The exact parameters (Yukawa coupling to the top quark $y_t$ as well as parameters from the Higgs potential and electroweak symmetry breaking) are less important than the huge prefactor $\Lambda^2$, which represents the upper limit of the theory's validity - worst case $10^{18}$ GeV (so-called UV range of the theory). The experimentally extracted renormalized mass $m_{H,r}$ of about $125$ GeV (so-called IR range of the theory) requires absorbing up to 16 orders of magnitude in the bare mass. The relatively tiny remaining value is perceived as unnaturally small and thus in need of explanation, as the widely separated energy scales must be artificially separated through the choice of the bare mass (not measurable). This does not happen for all other elementary particles because their loop corrections only contain terms of the form $\log (\Lambda)$ and thus limit their order of magnitude to $\mathcal{O}(10^1)$.

Moreover the following must be noted: UV sensitivity of the Higgs mass is only present in the sense that large terms from the UV sector appear in the corrections. This fact must be given attention as it represents at least a flaw for the concept of effective field theories (EFT). However, it does not mean that the Higgs mass could be sensitive to small changes in UV parameters (as would be feasible, for example, in changing initial conditions in dynamic systems) - these values are immutable.

Nevertheless, the problem of technical naturalness is not sharply separated from problems of accessing the world through effective theories; there is mutual influence. In recent literature, there has been a debate about the precise definition of technical naturalness and its influence on theory-building using EFTs. Giudice (2017, p.\,3) initially stated: "Naturalness is an inescapable consequence of the ingredients generally used to construct effective field theories." Hossenfelder (2017) then replied: "Of course it is not. If it was, why make it an additional requirement?" In a strict interpretation of Giudice's claim, an error can indeed be identified. Requiring technical naturalness is not equivalent to the requirement of decoupling energy scales, indispensable for the successful application of an EFT. Nevertheless, its violation raises suspicions about the universal applicability of EFTs. Williams (2019, p.\,17) attempts to reconcile by admitting: "(Technically) unnatural theories seem to violate the spirit, though not the letter, of the relationship between scales in effective field theory." Technical naturalness is thus a stricter principle. An EFT does not initially provide information about how large the quantum corrections $\Delta m_H$ is allowed to be. Values from the UV range do not yet contradict the use of EFTs. He explains: "Naturalness in this sense is motivated insofar as structural features of effective field theory lead us to expect a general insensitivity of low-energy observables to the structure of the theory at much higher energies, but it goes beyond what is strictly licensed by those structural features" (Williams 2019, ibid.). 

A clear distinction between decoupling and technical naturalness has already been attempted. Franklin (2020) differentiates between autonomy of microstates (IR dynamics is autonomous despite changes in the UV range) and autonomy of microlaws (IR dynamics is autonomous despite changes in UV parameters)\footnote{In a very similar way, Bain (2019) calls this difference heuristic and precise decoupling.}. However, for the lessons learned from the subsequent argumentation, it will be more helpful to differentiate between weak and strong autonomy of scales:
\\ \\
\textit{Weak autonomy of scales} guarantees that an effective theory $T^{eff}_i$, defined on an interval $[ L_i, \Lambda_i]$ (IR cutoff $L_i$ and UV cutoff $\Lambda_i$) empirically adequately describes its intended phenomena, i.e., on intervals $[0, L_i]$ and especially $[\Lambda_i, \infty)$, there are no crucial additional effects that could seriously influence the behavior of the phenomena. In other words, its violation indicates an explicit correspondence to distant scales that causally affect the phenomenon. This weak autonomy legitimizes the application of EFT. 
\\ \\
In the case of \textit{strong autonomy of scales}, contributions to phenomena in $T^{eff}_i$, expressed in units of $\Lambda_i$, are roughly of the order of $\mathcal{O}(10^1)$. However, their violation does not prevent the application of EFT due to the option of renormalization. The quantum corrections are only a crucial additional effect in the non-renormalized equations, which is then mathematically manageable. In other words, its violation hints an implicit correspondence to distant scales, that do not causally affect, however, the physical mass or any other quantity in the IR region.
\\ 

An analogy may help to digest the technical definition: A researcher is confronted with a comprehensive textbook on the knowledge of an entire discipline. They want to learn everything important about a specific phenomenon in this discipline and read the book effectively: In the table of contents they find that the topic is covered in Chapter 3, so they focus on this section, and ignore the rest of the book. Weak autonomy is violated if crucial information on the subject seems to be missing. One realizes that Chapter 3 offers a too narrow perspective to fully grasp the phenomenon - there must be additional input from the far-distant Chapter 28 that completes the picture. The researcher may have to make significant efforts and read the entire book in order to not miss anything. 

Strong autonomy is already violated if a footnote in Chapter 3 refers to some additional content in Chapter 28 (like huge quantum corrections giving a hint to the UV realm). The footnote contains all necessary information from Chapter 28, and effective reading is not hindered. The effective-reading enthusiast will suggest (artificially forcing far-distant content into the limited domain of the book): "Let's simply put the content of the footnote into the main text of Chapter 3, and everything will be fine." Indeed, the effective method remains usable and Chapter 28 can be left unread. The effective-reading sceptical (c.f. Athanasiou 2022 for the two personae) notes, however: "The phenomenon could have greater importance than initially assumed, could possibly occur universally and fundamentally, if there are correspondences to completely different sections of the book." This is, what broken strong and weak autonomy seem to have in common. But she has difficulties to convince the enthusiast - the breakdown of "autonomy of pages" attracts much more attention if one has to go beyond effective reading and explicitly experience the omnipresence (or fundamentality) of the phenomenon throughout the book.

With this illustration, it shall be clarified that even strong autonomy of scales is always closely intertwined with the application of effective theories. Both are expressions of the same (dubious) expectation to the world: that there should be no correspondence between UV and IR ranges. The violation of strong autonomy already questions whether effective theories provide the appropriate framework to adequately represent all phenomena occurring in them. In the case of the technically unnatural Higgs mass, an EFT remains usable; however, it takes this parameter as an artifact that would presumably need to be assigned to a different type of theories (fundamental). For this reason, aspects related to both strong and weak scale autonomy will be discussed in the following, and the critical examination of adherence to effective theories is inseparably linked to the revision of the hierarchy problem, too.

\subsection{How to Not Solve the Problem}

Several arguments cannot resolve this naturalness problem, as has already been highlighted by other authors:
\begin{itemize}
\item The calculation of corrections in dimensional regularization makes the cutoff disappear. \textit{Yes, but the naturalness problem will enter again through the backdoor, as worked out, e.g., by Williams (2015)\footnote{For the interested reader experienced in theoretical physics, there are more technical publications like (Harlander and Rosaler 2019a, 2019b) and (Rosaler 2021) discussing several naturalness formulations using a regulator-independent parametrization as well as various concepts of cutoffs and renormalization. We restrict ourselves to a non-technical, but philosophical (metatheoretical) perspective in this article, hopefully providing new insights at a different frontier.}}
\item Supersymmetry provides scalar loop corrections that could restore naturalness. \textit{Yes, but the crucial regions of the parameter space are already excluded. Unexcluded values for loop corrections may lower the value of the corrections, but the theory remains unnatural to some degree (Giudice 2017, Fischer 2024).}
\item Quantum gravity might not impose a limit on the Higgs boson, but be valid in isolation. The limit $\Lambda \to \infty$ would be justified, and the Higgs mass would be natural.\textit{ Yes, but the gauge coupling to hypercharge reaches its Landau pole around $10^{41}$ GeV. The theory remains effective (Craig 2022), unless progress in rigorous QFT solves the general problem of appearing Landau poles and provides a changed perspective on the hierarchy problem.}
\end{itemize}

Any argument based on the decoupling theorem (Appelquist, Carazzone 1975) must also be excluded. As expressed in several works (cf.\;Hartmann 2001), the proof relies on an outdated renormalization scheme and the assumption of the general renormalizability of theories, which has already been ruled out for quantum gravity.

\section{Limiting the Applicability of Technical Naturalness}
Confronted with the experimental value of the Higgs mass, one initially has several options to respond to the dependence of perturbative corrections on physics at extremely high energies that go beyond the explanations ruled out in Section 2.3: 
\begin{enumerate}
\item[a)] Physics beyond the standard model will some day provide a physical explanation for the light Higgs mass in some future \textit{(physical explanation)}.
\item[b)] Only the measured physical value of the Higgs mass has significance for the theory, being a constant of nature, and any discussion about its numerical value is unfruitful  \textit{(no explanation needed at all)}.
\item[c)] One needs to break out of internal physical reasoning and question the methodology of theoretical particle physics (especially the universal trust in effective field theories) itself \textit{(metatheoretical explanation)}.
\end{enumerate}
The BSM enthusiast in a) expects new physics providing a mechanism with explanatory power and is disappointed by the experiments of the past decades (Giudice 2017). We are not saying these experiments were in vain. If a single SM particle evades the otherwise reliable decoupling of low-energy scales from UV physics, this may have an explanation within particle physics. Temporarily ruling this out can and should be considered as a non-negotiable part of scientific practice. But for the pruposes of this paper, after decades of missing experimental evidence, let's suppose that no future BSM physics will be discovered that can restore the unnatural Higgs mass. 

The disappointed BSM physicist can simplify things and propose explanation b). The Higgs mass is what it is, and its numerical value should not be questioned at all: a prime example of a pseudo-problem. Hossenfelder (2018), for instance, brought this claim into a broader public stating that 125 GeV is simply a constant of nature, and the relevance of large quantum corrections is overestimated. Wetterich (1984, p.\;217) writes in one of the earliest analyses: "In any case, fine-tuning of bare parameters is not really the relevant problem (...) The relevant parameters are the physical parameters, since any predictions of a model must finally be expressed in terms of these". It is, therefore, unproblematic to compensate for the enormous correction terms $\propto \Lambda^2$ by an appropriate bare mass; in the end, the renormalized mass remains as the only isolated measurable quantity. Whether the correction terms have a dependence $\propto \log(\Lambda)$ or $\propto \Lambda^2$, there is no problematic fine-tuning, and the hierarchy problem loses significance. 

However, Wetterich's analysis might overlook a potential physical reality of quantum field theoretical corrections in $\Delta m_H$. The unique position of the Higgs boson as the only spin-0 particle in the described relation, which grants it a glimpse into UV physics by means of the large quantum corrections, has a physical significance - we might learn something from the large value of $\Delta m_H$, finally leading us to hypothesis $H_2$. The relation between spin-0 of the real particle and the virtual particles of the vacuum would give rise to a physical phenomenon that occurs in several cases of fundamental physics - the violation of autonomy of scales. Argueing in greater detail, why tautological statements (that its numerical value is what it is) are a too simple way out, would go clearly beyond the scope of this paper. So, in analogy to option a), let's also suppose that dismissing the hierarchy problem as a misguided debate about fixed constants of nature as a mere measurement result is not a satisfactory answer as well.

With these conditions, under which the paper is written - that options a) and b) will not solve the hierarchy problem - we can turn to a critical examination of the scope of autonomy of scales in fundamental physics, thereby elaborating reaction c). This has already happened in the recent past, albeit with sometimes drastic, consequences for physics and metaphysics. Wallace (2019), for instance, warns that the rejection of the principle of naturalness implies a rejection of reductionism as a overarching principle of foundational research. In this article, however, an opposite perspective - a positive consequence of rejecting the principle of naturalness - will be argued for, shedding an encouraging light on the naturalness debate about the Higgs mass (condensed in $H_2$). Reaction c) thereby has the diagnosis of a pseudo-problem in common with b). In c), however, it is caused by a dangerous mixture of metatheoretical and intratheoretical (physical) reasoning\footnote{This kind of pseudo-problem is not new: They have historically emerged, for instance, in efforts to explain the distances of planetary orbits from the Sun, where Kepler attempted to match these distances by nesting Platonic solids. Yet, these numerical values are not true anomalies. The issue arises from mistaking coincidental mathematical patterns that happened to emerge during the solar system’s formation for deep mathematical truths, similar to patterns that could appear around billions of other stars with entirely different configurations. Only this metatheoretical insight could reveal and eliminate the pseudo-problem.}.
\\ \\ 
Consider the impact of the UV-IR dialogue occurring in the context of the Higgs mass on the widely adopted methodology of EFTs. In the tower of quasi-autonomous domains in the sense of Cao and Schweber (1993), each described by an EFT, the existence of a scalar elementary particle (the Higgs) breaks this autonomy and establishes a dialogue between significantly different energy scales. Is it a coincidence that this phenomenon also appears in several mechanisms in the universe, which can be considered fundamental for good reasons? \textit{Ontological fundamentality}  expresses in this article that an entity, precisely a mathematical structure (be it the Higgs boson as a collection of quantum numbers or something entirely different), exists in a primary sense, that is, it does not depend on or cannot be traced back to other entities of more advanced (future) theories. For the Higgs boson in particular, we can dive into some details: First, a "naive" materialistic interpretation of \textit{fundamental} refers to the idea that a particle has no internal structure and is not composed of smaller components. The Higgs boson is regarded as a point-like, elementary particle within the framework of the standard model, without substructure or inner complexity (as for all other SM particles, however with no implications with respect to technical naturalness, as gauge and chiral symmetries keep the renormalized masses small). A second hint on fundamentality is the question of to what extent certain theoretical elements possess invariant characteristics under theory change. If a theoretical construct, such as the Higgs boson, consistently appears in a series of future series across different energy scales, this speaks to its robustness. For instance, Worrall (1989) or Williams (2019) interpret this robustness as an indication that such entities are not merely theoretical artifacts but rather reflect fundamental components of nature. A third and last perspective on fundamentality is a structural one and is directly connected to Worrall's work. Structure appears to be a more sophisticated, but more adequate way to characterize elementary particles: Quantum field theory shows that what we call "particles" are merely excitations or disturbances in underlying quantum fields, which seem to be fundamentally abstract mathematical structures. The materialistic view struggles to account for phenomena such as vacuum fluctuations, renormalization, and the creation-annihilation processes that defy classical intuition. These issues imply that the conventional notion of particles as concrete, self-contained objects doesn't capture the true nature of quantum entities. Instead, interpreting the Higgs field as mathematical structure better reflects how physical interactions emerge from deep, underlying symmetries and relationships rather than from independently existing material particles. Fundamentality enters the stage, if the mathematical structure of interest goes beyond a provisional construction, helpful to preliminarily describe an actually deeper phenomenon (e.g., the 4-fermion interaction as an empirically adequate description of the beta decay in earlier days, but revealed to be mathematically inconsistent and thus non-fundamental, being replaced by the GSW model containing W- and Z-bosons). We will discuss the consequences for $H_2$ in greater detail in Chapter 4.2.

These non-provisional mathematical structures seem to occur in various disciplines in fundamental  physics and are related with broken autonomy of scales: In three quite different mechanisms presented below, 1a) to 1c), such a delicate dependence of large length scales or small energy scales on the other end of the spectrum occurs. They have in common that physical explanations aiming to restore technical naturalness do not find application. On the other hand, the three successes of technical naturalness, 2a) to 2c), can be considered non-fundamental (as they are dealing with preliminary theories or composite entities). This comparison in 1) and 2) will provide the basis for hypothesis $H_1$. 
\begin{enumerate}
         \item[1a)]  One sensitivity arises already in the mesocosm itself, without the need to know anything about quantum effects: in the theory of complex systems. This represents a significant phenomenological extension of linear Newtonian physics without leaving classical physics. In linear mechanics, the dynamics of a system is causally unaffected by infinitesimal disturbances. The dynamics of nonlinear systems, on the other hand, show a sensitive dependence on the initial conditions with the slightest influence - small changes have a significant effect (UV-IR mixing). This is particularly evident in the butterfly effect, provocatively expressed in the question of whether the flapping of a butterfly's wings in Brazil can trigger a tornado in Texas (Lorenz 1972)\footnote{It may be argued that the Lorenz model, being derivable from the Navier-Stokes equations, might not represent a fundamental mechanism, weaking the key argument of this paper (the author wants to thank the reviewer for this issue). This perspective adopts a definition of fundamentality that emphasizes the basic equations of physics while overlooking the emergent structures and dynamics they produce. In many cases, core concepts in physics are not directly encoded in the fundamental equations, but rather emerge from them. This observation invites a broader view of what constitutes a fundamental aspect of a physical theory. For instance, Kellert (1993) emphasizes that chaos is not merely a special phenomenon in certain equations but instead exhibits universal patterns that appear in various physical systems. Such universalities strongly indicate that chaos is not merely a derived or contingent property but rather an intrinsic structure of dynamical systems, independent of their specific physical realization. This aligns, e.g., with the concept of the renormalization group theory in statistical physics: systems with different microscopic structures can exhibit the same universal critical phenomena on a macroscopic level. According to Kellert, chaos can unify different systems through universal patterns. This is a hallmark of fundamental concepts in physics: when a mechanism appears universally across many different theories, this suggests its fundamental significance. Kellert’s perspective thus demonstrates that derivability does not argue against the universality of a phenomenon.}. More precisely, the presence of turbulence in a multitude of complex systems, extremely dependent on tiny perturbations of initial conditions, hints that a manifestation of broken \textit{weak} autonomy of scales occurs as a fundamental property in the mesoscopic world as well, while the linear, i.e.\;effective theory describes only a small fraction of deterministic phenomena that show no dependencies on distant energy scales. \textit{Still, there are no discussions on physical explanations for broken autonomy of scales in nonlinear dynamics.}
         \item[1b)] A second UV-IR mixture arises under the assumption made for a theory of quantum gravity that space at the Planck length must have a non-commutative nature. This supplements the theory with the commutator relations between different operators, such as position and momentum, already necessary in quantum mechanics. To prevent gravitational collapse when localizing events with extreme precision ($l_{\mathrm{Planck}}$), one demands uncertainty conditions $\Delta x^\mu$ in the form $\Delta x^0(\Delta x^1+\Delta x^2+\Delta x^3) \geq l^2_{\mathrm{Planck}}$ and $\Delta x^1\Delta x^2+\Delta x^2\Delta x^3+\Delta x^3\Delta x^1 \geq l^2_{\mathrm{Planck}}$ in four dimensions. These uncertainties, whose causal effect is undoubted (violation of \textit{weak} autonomy of scales), are induced by the commutator $[x^\mu, x^\nu ]_{-} = i \theta^{\mu \nu}$ (Doplicher, Fredenhagen, Roberts 1995). In prose: If one wants to determine a coordinate with maximum precision (UV), a maximum uncertainty (IR) remains for the value of the other coordinate. This mechanism will probably represent a fundamental property of spacetime when quantum effects of gravity need to be considered.

UV-IR mixtures thus appear as a generic feature of non-commutative geometries, as pointed out by Minwalla, van Raamsdonk, and Seiberg (2000). Effective theories of gravity, as well as QFT, use the commutative limit $\theta \to 0$, in which this mixture does not occur. Koren (2020, p.\,265) illustrates as follows: "An effective field theorist living in a noncommutative space would have no way to understand the appearance of this infrared scale; its existence is intrinsically linked to the geometry of spacetime and to the far UV of the theory". \textit{Similarly, an EFT euphoric deals with the Higgs mass, but attempts to conceal a probably inherent fact with new physics. Surprisingly, UV-IR mixing is readily accepted when constructing a non-commutative spacetime.}
         \item[1c)] Thirdly, the specific properties of the quantum vacuum at the smallest length scales influence the expansion rate of the entire Universe (vacuum energy), which undoubtedly plays out on the largest available length scales when calculating the cosmological constant (see also Craig 2022 for an overview). Hence, \textit{weak}autonomy of scale is broken, the UV regime causally affects the IR regime. Unlike the balancing act performed between the electroweak scale ($10^{-18}$ m) and the Planck scale ($10^{-35}$ m) in the hierarchy problem of the Higgs mass, quantum effects in cosmology are even responsible for considerably larger length scales. On the other hand, effective theories usually separate cosmology and QFT, and the quest for dark energy remains unsolved - for its solution, the usage of EFTs potentially needs to be abandoned. Moreover, the violated \textit{strong} autonomy of scales exists in a way mathematically analogous to the Higgs mass: Summing the bare cosmological constant from general relativity and the value of the vacuum energy, $\tilde\lambda + \lambda=\Lambda$, results in handling a mismatch of 120 orders of magnitude between quantum effects and the observable value $\Lambda$ (to be distinguished from the cutoff).
     \end{enumerate}
The fundamental character of the phenomena from above is easily ascribed, be it the butterfly effect (an immutable principle in complex systems, also in refined future theories - see again Footnote 9), noncommutative geometry (fundamental structure of spacetime in theories of quantum gravity - "theories of everything") or the cosmological constant (determining the geometric evolution of the Universe itself). The crucial difference, explaining why the interpretation of broken autonomy of scale in case of the quantum field theoretical and cosmological hierarchy problems ($m_H$ and $\Lambda$) is way different than for the cases in 1a) to 1c), might be the distinction from Chapter 2: broken weak autonomy (in all three cases 1a-1c) and strong autonomy (hierarchy problems) - although they seem to have the implication of ontological fundamentality in common (detailed out in $H_2$)! Our book analogy may have helped to understand, why one did not problematize the UV-IR mixtures in 1a) to 1c) and dit not aim to cure them with artificial mechanisms ("new physics" in case of the technically unnatural Higgs mass), but instead accepted them as a "natural" consequence of fundamentality or universality. The second, more evident reason to speak about a hierarchy "problem" is the undoubted presence of the metatheoretical principle of technical naturalness in various cases. However, its applicability seems limited: Contrasted with violations of autonomy of scales in presumably fundamental phenomena, let us continue with three successful applications of the principle of technical naturalness in quantum field theory:\footnote{Instead of citing all single publications, we refer to (Williams 2015), where all three phenomena are explained in greater detail. It is helpful to remark, as Williams has already outlined, that only the last phenomenon was explained by explicitly applying the demand for naturalness. The former ones were recognized retrospectively as confirmations of the naturalness principle in QFT.}
\begin{enumerate}
         \item[2a)]  The divergent electron self-energy in classical electrodynamic, cured by the postulate of positrons by Dirac.
         \item[2b)] The pion mass difference, which can only be called technically natural, if the cutoff $\Lambda$ of its effective theory appears around 800 MeV, cured by the later discovered $\rho$ meson.
         \item[2c)]  The mass difference between the long- and short-lived state of the $K$ meson (kaon). Empirical reasons demand $\Lambda < 3$ GeV, which led to the prediction of the mass of the charm-quark
     \end{enumerate}
All three successes of technical naturalness share the commonality that the phenomena to which the principle was applied \textit{must not be considered fundamental, in stark contrast to the three cases previously discussed}. Instead, they involve composites (mesons) of elementary particles described in a narrow interval of the energy spectrum or misplace quantum fields (of the electron) into a classical field theory. Hence, the technical naturalness appears to be not universally applicable even within EFTs, but a metatheoretical guiding principle with a limited scope. From its successes we know, we derive that it can only be applied to non-fundamental phenomena modelled in provisional theories. This is the key message of hypothesis $H_1$. It wrong to conclude that technical naturalness applies simply because the standard model is an EFT and, simultaneously, the Higgs boson is an entity embedded into that EFT. We also need to know something about the ontological status of this entity.

 It is unclear why the decoupling of distant energy scales is a widespread phenomenon that simplifies life for physicists and we do not want to downplay the obvious empirical success achieved through the application of EFTs and the naturalness principle. In addition to the mentioned cases from QFT, this decoupling holds equally true in classical physics. It is the reason why physics could gradually approach more challenging theories, and the description of an apple falling from a tree did not require quantum gravity. It is the \textit{conditio sine qua non} of an effective theory. However, there is no reason to assume that autonomy of scales must be preserved outside of effective theories. Quite the opposite: An effective theory aims to provisionally and simplifiedly represent a part of reality (a section of the entire energy scale) - a fundamental theory, on the other hand, captures elements of reality itself (not limited to specific energy ranges). A violation of autonomy of scales, in both the weak and strong sense, seems almost a \textit{natural} consequence! It must be emphasized that our viewpoint is far from consensus: Crowther (2019), for instance, provides a list of criteria a fundamental theory shall fulfil and explicitly mentions naturalness as one criterion. However, based on the comparison from above giving rise to hypothesis $H_1$, which arguments can be provided that the metatheoretical principle of naturalness has to be fulfilled beyond the end of the chain of EFTs - to be fulfilled in final theories describing fundamental phenomena as well? Giudice (2017, p.\,8) writes "one way or another, naturalness will still play a role in the post-naturalness era". Of course, we agree: it will certainly play some role. But like other advocates of naturalness, he does not concretize the scope of applicability. This is where $H_1$ enters the stage and adds a first new aspect to the ongoing naturalness debate. It restricts the applicability of technical naturalness to non-fundamental entities and relations within EFTs. Presumably, technical naturalness will play a role in future effective theories while predicting new physics as Giudice (2017) already pointed out. But the precise formulation of this thesis is crucial - via an ontological vocabulary rather than the commonly used, too narrow vocabulary of EFTs, which is closely related to it: Naturalness is of course tied to the concept of EFTs by definition. In contrast to the formulation presented here, however, the Higgs (embedded in the EFT of the standard model) inevitably raises a naturalness problem. The Higgs boson can only be clearly separated from the phenomena explained in 2a to 2c in the following chapter by arguing about its status of ontological fundamentality.

 Arguing for the possibility of an ontologically fundamental status of the Higgs boson within EFTs in the next chapter will reveal the hierarchy problem of the Higgs mass as a pseudo-problem, thereby deducing $H_2$ from $H_1$.

\section{Placing the Higgs Outside of EFTs}
The absence of technical naturalness becomes a cause for concern only under the assumption discussed before and consolidated in $H_1$: that the seemingly reductionist theory sequence is endless and the series of effective theories does not terminate. Lacking autonomy of scales would then indeed signal the failure of reductionism.  In this case, against whose plausibility we will now present arguments,  the general modus operandi of foundational physics would be challenged to such an extent that missing autonomy of scales must be elevated to the status of a naturalness \textit{problem}, rather than characterizing it as an almost desirable trait, possibly revealing a fundamental character of the entity of concern. We will see below how influential publications have cemented the strict adherence to EFTs, thereby hindering a comprehensive consideration of the hierarchy problem. But what is there to be said against the omnipresence of EFT and, over and above, for the fundamental nature of the Higgs boson?

\subsection{The World is More Than a Set of Quasi-Autonomous Domains.}
In the following discussion, we critically evaluate the ontological consequences of considering autonomy of scales. To strengthen the argument for the reevaluation of the absence of technical naturalness, we turn to a widely discussed work by Cao and Schweber (1993), which describes physics as the aforementioned infinite tower of quasi-autonomous domains, each ontologically equivalent and effectively described. Thus, the authors not only adopt an anti-reductionist and anti-fundamentalist stance (\textit{infinite} tower), but they also postulate an ontological pluralism, thereby inadequately representing the methods and understanding of physics in the last century, as we will see below. The consequences of that world view have been criticized extensively (e.g., Hartmann 2001), partly due to the use of the decoupling theorem (Appelquist, Carazzone 1975), which is today excluded in naturalness debate for good reasons beforehand. One should of course admit that a level-by-level reductionism—where one theory reduces to another without requiring a single lowest level—could still be compatible with this framework. Nevertheless, we will once again challenge the worldview outlined there, as the scope of the work has influenced the assessment of absent autonomy of scales. Thereby, it seems, the dogmatic adherence\footnote{This term refers to the overgeneralisation of the naturalness principle - the unjustified extrapolation into the realm of phenomena that do not strictly belong to EFTs.} to EFTs ultimately makes the hierarchy problem of the Higgs mass so persistent in the philosophical and particle physics debate. So let us liberate ourselves from the dogma that effective theories are the only way we can describe the universe.
 
To do that, it is helpful in a first step to acknowledge that the coexistence of different EFTs by no means allows the conclusion that our world is ontologically plural. An effective theory, by definition, cannot enable ontological determinations. Moreover, no actual autonomy of these respective domains can be established. There are continuous transitions in, for example, the theory of strong nuclear force, where the approximate validity of point-like nucleons is steadily reduced, and at no discrete point does the absolute necessity of quarks become apparent. Such smooth transitions are already present in the construction of the theory to be effective. The transition to another domain is always at the discretion of the researcher, who decides what deviations are tolerable for the given context.

Let us recap the modus operandi of a foundational physicist: She deems a theory effective because she understands it as a provisional approximation to a cosmos governed by a fundamental theory. At the time of developing the EFT, this fundamental theory is evidently not within intellectual and perhaps empirical reach yet, so one settles for an approximation reliable in certain domains. During the convergence (in its empirical adequacy) of the sequence of theories reliable in certain domains against a fundamental theory reliable in all domains, it represents a precursor of merely epistemic value. Instead of making ontologically fundamental statements already, it simply propels the physicist a significant step forward on this path. This represents a central assumption of foundational physics, abandoning it would instead deny the epistemic value of such interim successes. This pragmatic approach to the microcosm, whose exploration is solely driven by the pure desire for knowledge and cannot be motivated by practical significance, is legitimized only by the attainability of a foundation. EFTs thus serve as instruments to approach fundamental theories. This approximation is missing in the infinitely layered tower of EFTs. The potential senselessness in the quest for fundamental physics in this picture is not an argument in itself that can be used against Cao and Schweber. It is rather the fact that ever higher precision and ever smaller distances to the extreme scales of physics (Planck scale) do not give any reason to believe that we are just as far from the foundation of the tower as in Newtonian mechanics, as suggested by the image of the infinite tower. In short, Cao and Schweber paint a false picture of the way of working of a foundational physicist in exploring the world, as well as of the world itself. We are certainly able to reach at least some of the fundamental elements of being, allowing for ontological commitments. Ontological plurality rules out the existence of fundamental entities, the search for which is actually one of the main tasks of fundamental physics. Pure EFTs, on the other hand, are constantly being replaced by more and more glances on the foundation of our cosmos, where the ‘tower of EFTs’ gradually may end and is replaced.

With these arguments, the long-term significance of EFTs and thus the adherence to technical naturalness is once again questioned, especially questioning it as a guiding principle for "final" theories to come. Williams (2015) concludes that its violation in the case of the Higgs mass is a counterargument to the worldview previously presented by Cao and Schweber (1993). However, even its general consideration (without knowing anything about autonomy of scales) reveals, as shown, sufficient incoherence with the discipline of physics to reject it. Hence, let us move beyond purely effective theories and incorporate fundamental entities into them.

\subsection{Embedding a Fundamental Higgs into Preliminary EFTs }

Athanasiou (2022, p.\,115) categorizes the work of Cao and Schweber (1993) into that of        EFT-enthusiasts (among optimists, cautious and sceptics). In contrast, when addressing EFT-sceptics, he states, "Clearly, the sceptical road is the path of the patient: in the absence of a completely satisfactory theory, thou shalt not make any ontological commitment". Indeed, this commandment brings us back to the initial question: Has the last word already been spoken concerning the real, the true nature of the Higgs boson? Only in that case we had a valid argument that the absence of technical naturalness in its renormalized mass can be interpreted as a sign of having reached final, scale-crossing theory without causing any "naturalness problem" (the essence of $H_2$, mainly deduced from $H_1$) - as for noncommutative spacetime or senstive dependence on initial conditions of complex systems (cf.\;Chapter 3). Koren (2020) expects the Higgs mass to be a parameter resulting from input values of a UV theory in low-energy theory, even if it is unknown whether the Higgs mass ultimately emerges as part of a larger multiplet, as a bound state of fermions, or as an excitation in string theory. However, as we have already noted, those theories beyond the standard model, especially string theory, remain experimentally unconfirmed and pure speculation. There is no convincing reason to believe that the Higgs boson might not be a ontologically fundamental entity. Thus, it is hoped that the research program of BSM physics will ultimately lead to the desired situation, allowing for actual ontological commitments about the smallest building blocks of matter for the first time. 
\\
However, how can the presumably fundamental nature of the Higgs field in the universe - the crucial assumption for $H_2$ - be reconciled with the fact that the standard model as a whole must be assigned to the class of EFTs, making final ontological conclusions premature? The answer to this last crucial question, which ultimately enables our envisaged diagnosis of a pseudo-problem, will be provided in the following by the approach of Worrall (1989), among others. Worrall’s structural realism emphasizes that it is primarily the abstract mathematical structures that prove resilient through theory change. A classic example of this is the transition from Newton’s theory of gravitation to Einstein’s general theory of relativity. While Newton’s approach describes gravity as a force between two masses that diminishes with the square of the distance, Einstein interprets gravity as the curvature of spacetime. Despite these fundamental conceptual differences, the characteristic $1/r^2$  dependency is preserved in Einstein’s theory as an approximation in the weak-field limit, when considering the gravitational potential. This persistence is no coincidence but reflects deeply rooted geometric principles. Newton’s heuristic postulation of this mathematical structure is derived by Einstein as a fundamental feature with a geometric justification. In this sense, structural realism offers a compelling perspective on how knowledge about nature can remain robust and enduring despite shifts in theoretical frameworks.

In line with Worrall, one argues that the Higgs boson is not merely to be understood as a particle within the standard model, but as an expression of a fundamental mathematical structure that persists even in theories beyond the standard model. Even if future models replace the standard model and introduce different dynamics or additional fields, the fundamental role of the Higgs boson as a mediator of the mechanism of spontaneous symmetry breaking would likely remain. This structural persistence—the recurring mathematical relation between fields that generates mass—reflects a more robust truth than the materialist ontology outlined in Chapter 3 ever could. If a theoretical construct, such as the Higgs boson, consistently appears across different energy scales and within various models, this speaks to its robustness. One can interpret this robustness as an indication that such entities are not merely theoretical artifacts but rather reflect fundamental components of nature.

Given the evident limitations of the standard model to specific energy scales (an EFT by construction), labelling the entire model as a fundamental theory of particles is unjustified.   Whether the future replacement of the theory as a whole is known in advance or not, Worrall would argue that elements interpretable as ontologically fundamental were causally linked to the temporary and future empirical adequacy of the theory. One attributes lasting value to them for theory-building and sees them as guarantors of success for future theories. 
If such element of a theory breaks autonomy of scales (as in the case of the Higgs mass), the  pure approach via EFTs (as the standard model as a whole), gradually comes to an end by $H_2$, if no physical explanations for this behavior can be demonstrated. However, failure within the framework of such pragmatic theories need not be considered a problem, as the term "naturalness problem" has always suggested: Keeping in mind the "natural" occurrence of UV-IR-mixtures in the examples from Chapter 3 ($H_1$) and simultaneously respecting th success of the theoretical description of the world using Higgs bosons up to very high energy scales, one should instead attribute to the Higgs boson a fundamental character and cease the dogmatic attempt to artificially force a fundamental phenomenon into the scope of provisional theories. This is ultimately the justification of $H_2$. Violations of strong autonomy of scales may thus be used as an encouragement to replace the labelling of the standard model to be able only to answer penultimate questions by a more subtle perspective: There may be entities such as the Higgs boson that lie completely outside the stack of effective theories as proclaimed by Cao and Schweber (1993) and remain unaffected by broken scale autonomy. They are not only empirically adequate on limited scales, but will be reflected in the complete picture of our cosmos. These fixed points in the dynamic system of changing theories will ultimately be the constituents of a fundamental inventory of our cosmos. Only this nuanced view of the overall provisional standard model can ultimately enable the main thesis of the paper. 

In this view, it sends a positive signal to all the physicists searching for fundamental entities: It is highly plausible that future theories will employ the same mechanism turning the Higgs boson into an entity that will survive any theory change to come\footnote{One last, but crucial distinction shall be noted: The fermions of the standard model can still be considered fundamental and irreducible, even though their quantum corrections to their mass are natural through chiral symmetries ('t Hooft natural). As long as there are no signs of substructures of these assumed point-like and elementary particles, Worralls arguments can also be applied here. Thus, violated strong autonomy of scales is not a necessary criterion for attributing (approximate) truth to theoretical assumptions. It must be rather understood as a sufficient criterion once the fundamental nature of the phenomenon has been sufficiently made plausible.}

As a perspective for a future investigation and challenge to the application of Worrall's concept to QFT in general, we note that the lack of mathematical understanding of QFT (missing axiomatic constructability, Haag's theorem, Landau poles, divergent perturbative series, ...) carries the risk that its use in present-day particle physics could be fundamentally doomed. Grinbaum and Rivat (2020, p.\,11) remark: "If the formalism of the new theory after a future theory change happens to be incommensurable with the mathematical structure of QFT, even the parts favored by selective realists will become void of ontological status". The Higgs boson with its UV-sensitive mass corrections is currently embedded in the formalism of QFT, too. We cannot exclude that a reformed version of QFT could shed new light on the problem of technical naturalness in its mathematical definition over $\Lambda^2$. The role of rigorous mathematics for an (approximately) true description of the world must therefore be thoroughly examined\footnote{Perhaps, to make ontological statements about the world, one cannot settle for effective or pragmatic mathematics. In other words, the question arises: To what extent must fundamental (non-pragmatic, not inconsistency-ignoring) mathematics of an axiomatic QFT and the description of fundamental entities go hand in hand for the theory to allow ontological statements? The mathematics used today to describe the (nevertheless, in our assumption, fundamental) Higgs boson, for example, could distort its true nature to such an extent that present-day physicists are not yet privy to any truth. Such distortion could occur, for example, in resolving the previously described Landau pole, which the gauge coupling to the hypercharge reaches at finite energies, prohibiting the limit $\Lambda \to \infty$. The residual jeopardy of the fundamental character of the Higgs boson in the current state of particle physics could be eliminated by the mathematics of the future, but also delay the claim of fundamentality in any sense of the word explained in Chapter 3.}.

\section{Conclusion}

This article provided novel (metatheoretical) arguments that the restoration of technical naturalness of the Higgs mass with new physics represents an engagement with a pseudo-problem under a crucial assumption. However, reinterpreting the failure of the guiding principle of technical naturalness in case of the Higgs boson allowed for ontological commitments about high-energy physics: a statement about its status as a fundamental entity of our cosmos - embedded into an effective field theory (the standard model) being by definition a provisional construction. A final justifaction came from applying the concept of Worrall's (1989) and Williams' (2019) robustness of certain theoretical entities to the standard model, claiming that there are already fundamental  entities and relations that will survive any future theory change, included into that overall effective framework. Hence, the smallness of the Higgs mass reveals itself as a glimmer of hope: A hint at the fundamental nature of the affected phenomenon, which to reveal is the long-term mission of fundamental physics. We believe that as long as one only dealt with the artificial restoration of strong autonomy of scales in the corrections to the Higgs mass through theories without empirical evidence, there is a degenerating element in the BSM research program.

 This article denied the status of a problem while simultaneously enabling a progressive shift. In this shift, the limit of the applicability of pure EFTs is accepted, legitimizing greater openness in theory-building and evaluation without attacking the general approach of BSM physics through effective theories: What impact does the discussion about the gradual dissolving (purely) effective theories have on the near future of BSM physics in general? Regarding the forthcoming use of EFTs, we can only align ourself with the pragmatism of Georgi (1993, p.\,215) writing: "In addition to being a great convenience, effective field theory allows us to ask all the really scientific questions that we want to ask without committing ourselves to a picture of what happens at arbitrarily high energy". Indeed, the broken autonomy of scales in the correction to the Higgs mass should be understood as a failure of the adherence to pure EFTs for this specific scenario. The survival of invariants under future theory changes, indicators for fundamentality, provides a compromise without being overly optimistic about the persistence of the standard model as a whole. The more general use of EFTs remains a central prerequisite for a thriving particle physics in the future. This article merely criticized the dogmatic adherence to principles for which there are convincing reasons to release them: The simple and, given the current empirical evidence, plausible assumption of a Higgs boson that is ontologically fundamental, suffices (with $H_1$ and $H_2$) to turn the hierarchy problem into a pseudo-problem. 

Ultimately, it seems that the question of the unnatural Higgs mass must be asked at a deeper level. Instead of asking ‘what physical mechanism allows for a small Higgs mass?’, the crucial question is actually a metatheoretical one: ‘is there an ontological status that allows for a small Higgs mass without requiring a physical mechanism?’. The latter question can be answered in the affirmative under the assumption of ontological fundamentality, the former becomes superfluous.
 \subsection*{Acknowledgements}
We thank Ulrich Krohs, Raimar Wulkenhaar, Michael te Vrugt und Florian Fabry for valuable discussions and all members of the colloquium for the discussion of publications of the Philosophisches Seminar (University of Münster) for a critical review.  Furthermore, we would like to express our gratitude for the detailed and thoughtful critiques of the manuscript by the reviewer.

\newpage
\noindent \textbf{Bibliography} \\ \\ 
\begin{small}
\begin{onehalfspace}
Adlam, E.: \textit{Fundamental?} In: Aguirre, A., Merali, Z.\,(Ed.), Springer (2019)  \\  
Appelquist, T. and Carazzone, J.: \textit{Infrared singularities and massive fields}. \url{https://doi.org/10.1103/PhysRevD.11.2856}. Phys.\,Rev.\,D, 11(10): 2856–2861 (1975) \\  
Bain, J.: \textit{Why Be Natural?}. \url{https://doi.org/10.1007/s10701-019-00249-z}. Found. Phys. 49(9):  898–914 (2019) \\ 
Banks, T.: \textit{Modern Quantum Field Theory.} Cambridge University Press, Cambridge (2008)\\
Borrelli, A., Castellani, E.: \textit{The Practice of Naturalness: A Historical-Philosophical Perspective}. \url{https://doi.org/10.1007/s10701-019-00287-7}. Found. Phys. 49(1): 860-878 (2019) \\
Cao, T.Y., Schweber, S.: \textit{The conceptual foundations and the philosophical aspects of renormalization theory}. \url{https://doi.org/10.1007/BF01255832}. Synthese 97(1): 33-108 (1993) \\
Craig, N.: \textit{Naturalness. A Snowmass White Paper}. Submitted to the Proceedings of the US Community Study on the Future of Particle Physics. \url{https://doi.org/10.48550/arXiv.2205.05708}. arXiv:2205.05708 (2021) \\ 
Crowther, K.: \textit{When Do We Stop Digging? Conditions on a Fundamental Theory of Physics}. In: Aguirre, A., Merali, Z.\,(Ed.), Springer (2019) \\
Dine, M.: \textit{Naturalness Under Stress Annual Review of Nuclear and Particle Science}. \url{https://doi.org/10.48550/arXiv.1501.01035}
arXiv:1501.01035 [hep-ph] (2015) \\ 
Dirac, P.: \textit{Cosmological Models and the Large Numbers Hypothesis}. Proceedings of the Royal Society of London A. 338 (1615)X 439–446 (1974) \\ 
 Doplicher, S., Fredenhagen, K., Roberts, J.E.. \textit{The Quantum structure of spacetime
at the Planck scale and quantum fields}. \url{https://doi.org/10.1007/BF02104515}. Commun. Math. Phys. 172: 187–220,
(1995) hep-th/0303037. \\
Fischer, E.: \textit{Naturalness and the Forward-Looking Justification of Scientific Principles}. \url{https://doi.org/10.1017/psa.2023.5}. Philosophy of Science 90(5): 1050–1059 (2023) \\ 
Fischer, E.: \textit{The Promise of Supersymmetry}. Synthese 203(6) :1-21 (2024) \\ 
Franklin, A.: \textit{Whence the Effectiveness of Effective Field Theories?} \url{https://doi.org/10.1093/bjps/axy050}. British
Journal for the Philosophy of Science 71(4): 1235–1259 (2020) \\ 
Gaillard, M.K., Lee, B.W.: \textit{Rare decay modes of the K mesons in gauge theories.} \url{https://doi.org/10.1103/PhysRevD.10.897}. Phys.\,Rev.\,D,
10(3), 897–916 (1974) \\ 
Gaillard, M.K., Lee, B.W.: \textit{Search for charm.} \url{https://doi.org/10.1103/RevModPhys.47.277}.  Reviews of Modern Physics, 47(2), 277–310 (1975) \\ 
Gell-Mann, M.: \textit{The Eightfold Way: A Theory of Strong Interaction Symmetry} (Report). \url{https://doi.org/10.2172/4008239}. California Institute of Technology Synchrotron Laboratory (1961) \\ 
Georgi, H.: \textit{Effective Field Theory}. \url{https://doi.org/10.1146/annurev.ns.43.120193.001233}. Annual Review of Nuclear and Particle
Science 43: 209–252 (1993) \\ 
Giudice, G.F.: \textit{Naturally speaking: The naturalness criterion and physics and LHC.} \url{https://doi.org/10.1142/9789812779762_0010}. arXiv:0801.2562 (2008) \\ 
Giudice, G.F.: \textit{The Dawn of the Post-Naturalness Era}. \url{
https://doi.org/10.48550/arXiv.1710.07663}. Contribution to the volume \textit{From My Vast
Repertoire - The Legacy of Guido Altarelli}. arXiv:1710.07663 (2017) \\ 
Grinbaum, A., Rivat, S: \textit{Philosophical foundations of effective field theories}. \url{https://doi.org/10.1140/epja/s10050-020-00089-w}. Eur.\,Phys.\,J.\,A 56: 90 (2020) \\ 
Harlander, R., Rosaler, J.: \textit{Naturalness, Wilsonian renormalization, and “fundamental parameters” in quantum field theory}. \url{https://doi.org/10.1016/j.shpsb.2018.12.003}. Studies in History and Philosophy of Science B 66: 118-134 (2019) \\ 
Harlander, R., Rosaler, J.: \textit{Higgs Naturalness and Renormalized Parameters}. \url{https://doi.org/10.1007/s10701-019-00296-6}. Foundations of Physics 49(9): 879-897 (2019) \\
Hossenfelder, S.: \textit{Naturalness is dead. Long live naturalness}. Article in backreaction.blogspot.com (2017) \\ 
Hossenfelder, S.: \textit{Das hässliche Universum.} S. Fischer Verlag (2018)\\
Hossenfelder, S.: \textit{Screams for Explanation: Finetuning and Naturalness in the Foundations of Physics}. \url{
https://doi.org/10.1007/s11229-019-02377-5}. Synthese 198: 3727–3745 (2021) \\ 
Kellert, S. H.; \textit{In the Wake of Chaos}. Chicago, IL: University of Chicago Press. \url{https://doi.org/10.7208/9780226429823}. (1993) \\
Koren, S.: \textit{The Hierarchy Problem: From the Fundamentals to the Frontiers}. Dissertation University of California, Santa Barbara (2020) \\ 
Kragh, H.: \textit{Dirac: A Scientific Biography}. Cambridge University Press (1990) \\ 
Lorenz, E.N.: \textit{Predictability: Does the flap of a butterfly's wings in Brazil set off a tornado in Texas?} Title of a talk from 1972 during Annual Conference of the American Association for the Advancement of Science. In: Science 320: 431 (2008) \\ 
 Minwalla, S., van Raamsdonk, M., Seiberg, N.: \textit{Noncommutative perturbative dynamics}. \url{
https://doi.org/10.1088/1126-6708/2000/02/020}. JHEP, 02: 020 (2000)\\ 
Penrose, R.: \textit{The role of aesthetics in pure and applied mathematics.} Bulletin of the Institute of Mathematics and its Applications, 10, 266–271 (1974) \\ 
Rosaler, J.: \textit{Dogmas of Effective Field Theory: Scheme Dependence, Fundamental Parameters, and the Many Faces of the Higgs Naturalness Principle}. \url{https://doi.org/10.1007/s10701-021-00510-4}. Found. Phys. 52(1): 1-32 (2021)\\ 
Susskind, L.: \textit{Dynamics of spontaneous symmetry breaking in the Weinberg-Salam theory}. \url{https://doi.org/10.1103/PhysRevD.20.2619}. Phys.\,Rev.\,D
20, 2619 (1970) \\ 
 Tegmark, M.: \textit{The mathematical universe}. \url{https://doi.org/10.1007/s10701-007-9186-9}. Foundations of Physics 38: 101–150 (2008) \\
't Hooft, G.: \textit{Naturalness, Chiral Symmetry, and Spontaneous Symmetry Breaking}. \url{https://doi.org/10.1007/978-1-4684-7571-5_9}. Proc.\,of
1979 Cargèse Institute on Recent Developments in Gauge Theories, New York, Plenum Press (1980)\\ 
Wallace, D.: \textit{Naturalness and Emergence}. \url{https://doi.org/10.1093/monist/onz022}. The Monist 102(4): 499–524 (2019) \\  
Weinberg, S: \textit{Approximate Symmetries and Pseudo-Goldstone Bosons}. \url{https://doi.org/10.1103/PhysRevLett.29.1698}. Phys.\,Rev.\,Lett. 29 (1972) \\
Wells, D.W.: \textit{The Utility of Naturalness, and how its Application to Quantum Electrodynamics envisages
the standard model and Higgs Boson}. \url{https://doi.org/10.1016/j.shpsb.2015.01.002}. Studies in History and Philosophy of Modern Physics 49: 102–108 (2013) \\  
Wetterich, C.: \textit{Fine-tuning problem and the renormalization group}. \url{https://doi.org/10.1016/0370-2693(84)90923-7}. Physics Letters
B 140.3, 215–222 (1984)  \\
Williams, P.: \textit{Naturalness, the autonomy of scales, and the 125 GeV Higgs.} \url{https://doi.org/10.1016/j.shpsb.2015.05.003}. Studies in History and
Philosophy of Science Part B: Studies in History and Philosophy of Modern Physics 51: 82-96 (2015) \\
Williams, P.: \textit{Scientific Realism Made Effective}. \url{https://doi.org/10.1093/bjps/axx043}. British Journal for the Philosophy of Science 70(1): 209-237 (2019) \\ 
\end{onehalfspace}
 \end{small}





\end{document}